**Spatial Charge Inhomogeneity and Defect States in Topological Dirac Semimetal Thin Films**

Mark T. Edmonds[1,2†*], James L. Collins[1,2†], Jack Hellerstedt[1,2], Indra Yudhistira[3], Lídia C. Gomes[3], João N. B. Rodrigues[3], Shaffique Adam[3,4], Michael S. Fuhrer[1,2*]

1. School of Physics and Astronomy, Monash University, Clayton VIC 3800, Australia
2. Monash Centre for Atomically Thin Materials, Monash University, Clayton VIC 3800 Australia
3. Department of Physics and Centre for Advanced 2D Materials, National University of Singapore, 117551, Singapore
4. Yale-NUS College, 6 College Avenue East, 138614, Singapore

† These authors contributed equally to this work
* Corresponding author: mark.edmonds@monash.edu and michael.fuhrer@monash.edu

**The close approach of the Fermi energy $E_F$ of a Dirac semimetal to the Dirac point $E_D$ uncovers new physics such as velocity renormalization,[1,2,3] and the Dirac plasma [4,5] at $|E_F - E_D| < k_B T$, where $k_B T$ is the thermal energy. In graphene, substrate disorder drives fluctuations in $E_F$. Three-dimensional topological Dirac semimetals (TDS)[6,7] obviate the substrate, and should show reduced $E_F$ fluctuations due to better metallic screening and higher dielectric constants. Here we map the potential fluctuations in TDS $Na_3Bi$ using a scanning tunneling microscope. The rms potential fluctuations are significantly smaller than room temperature ($\Delta E_{F,rms}$ = 4-6 meV = 40-70 K) and comparable to the highest quality graphene on h-BN;[8] far smaller than graphene on $SiO_2$,[9,10] or the Dirac surface state of a topological insulator.[11] Surface Na vacancies produce a novel resonance close to the Dirac point with surprisingly large spatial extent and provides a unique way to tune the surface density of states in a TDS thin-film material.**

Three-dimensional topological Dirac semimetals (TDS) such as $Na_3Bi$ and $Cd_3As_2$[7,12,13] express the pseudorelativistic physics of two-dimensional Dirac material graphene, but extended to three dimensions. TDS can yield ultra-high mobilities[14] as well as new physics such as the chiral anomaly.[15] As the Fermi energy $E_F$ measured relative to the Dirac point energy $E_D$ approaches zero, the carrier density tends to zero and metallic screening disappears. Poorly screened disorder induces spatial fluctuations in the local $E_D$ and hence fluctuations $\Delta E_F$, corresponding to local electron and hole "puddles". The carriers in puddles in turn screen the disorder, with $\Delta E_F$ determined self-consistently by the disorder and the screening properties of the Dirac materials.[16,17] Puddles have been visualized in graphene using scanning single electron transistor microscopy[10] and scanning tunnelling spectroscopy,[8,9,18] where the fluctuations are largely governed by the underlying substrate.

Here we use scanning tunneling microscopy and spectroscopy (STM/STS) to probe $\Delta E_F$ fluctuations in 20 nm $Na_3Bi$ films on both semiconducting [Si(111)] and insulating [$\alpha$-$Al_2O_3$(0001)] substrates. Figure 1(a) shows a large area (400 nm x 380 nm) topographic STM image of a thin 20nm film of $Na_3Bi$ on Si(111), with several atomically flat terraces >100 nm in size. The inset to Figure 1(a) reveals the (1 x 1) Na-terminated atomic lattice (a = 5.45 Å) with an individual Na vacancy. Figures 1(b) (45 nm x 45 nm taken immediately after growth) and 1(c) (30nm x 30nm taken seven days after growth) show the topography of two atomically flat regions away from step edges or the screw dislocations seen in Fig. 1(a). Figure 1(d) shows an atomically flat region (30 nm x 30 nm) of $Na_3Bi$ grown on sapphire ($\alpha$-$Al_2O_3$(0001). Whilst atomically flat regions of Na$_3$Bi up to 100 nm x 100 nm can be obtained on Si(111), sparse defect cluster sites (few per 100 nm x 100 nm) give rise to tip-induced ionization ring features[21,22] that will be discussed further below. This necessitated focusing on smaller areas free of ionization rings in order to unambiguously determine the variation in Dirac point.

Figure 1(e) shows area-averaged scanning tunneling spectra (STS) of the Na$_3$Bi film. STS measures the differential conductance d$I$/d$V$ as a function of sample bias $V$, proportional to the local density of states (DOS) at energy $eV$ relative to $E_F$. STS was performed at points on a grid encompassing the entire regions shown in Figures 1(b & c), along with two other regions not shown. Figure 1(e) shows the averaged spectra of these four areas. Two key features are prominent in all spectra: a distinct minimum in the DOS corresponding to the Dirac-point energy ($E_D$), and a resonant feature ~35 meV below $E_D$, labelled D. As-grown, the Dirac point is located ~20 meV above the Fermi level indicating $p$-type doping consistent with angle-resolved photoelectron spectroscopy (ARPES) measurements.[23] Seven days after growth, the Dirac point has shifted to approximately 15 meV below the Fermi level, reflecting a gradual global $n$-type doping of the Na$_3$Bi due to the adsorption of atomic species present in UHV. This adsorption results in the formation of intermittent impurity clusters on the surface, as such all topography and spectroscopic measurements were deliberately performed away from such sites. The resonance feature D is unambiguously tied to $E_D$, and not to the $E_F$, as the relative energy shift of D with respect to $E_D$ remains unchanged within experimental accuracy during the transition from $p$-type to $n$-type doping.

We first turn to the spatial variation of the Dirac point energy, $E_D$, found by tracking the position of the minimum differential conductance in STS (alternatively, the shift in the defect resonance D in each spectrum gives similar results). Figures 2(a) and 2(b) show the spatial variation of $E_D$ for regions 'A' and 'B' corresponding to Figures 1(b) and 1(c) respectively. In both Dirac-point energy maps, it can be seen that a clear, continuously connected local potential modulation emerges, correlated on a scale much larger than the crystal lattice or point-spectroscopy grid. This modulation in $E_D$ represents the puddling of charge density at the surface. The variation in $E_D$ is found to be correlated with the individual Na vacancy sites which possess a more positive Dirac point energy, identifying the Na vacancies as acceptors and a significant source of charge disorder (see below and Supplementary Information). In addition, Figure 2(c) shows the local $E_D$ of 20 nm Na$_3$Bi on α-Al$_2$O$_3$(0001) (labelled Region C) which possesses a larger $n$-type doping. The upper, middle and lower panels of Fig. 2(d) show histograms of $E_D$ relative to $E_F$ for the scans in Figs. 2(a)-(c) respectively. From spatial autocorrelation analysis of the puddle maps, we determine spatial coherence lengths ξ of 13.4 ± 5.2 nm, 9.3 ± 2.4 nm and 5.1 ± 1.9 nm for Regions A, B and C respectively (see SI for calculation details). The observed histograms of $E_D$ have mean values 19.7 ± 1.7 meV (Region A), -15.4 ± 1.3 meV (Region B) and -47.2 ± 0.6 meV (Region C), and standard deviations $\Delta E_F$ of 5.6 meV, 4.2 meV and 3.5 meV respectively. These values are comparable to the 5.4 meV observed for graphene on h-BN.[8] However, undersampling results in an underestimate in the magnitude of $\Delta E_F$ by a factor $(L/\xi)^2/[(L/\xi)^2 -1]$ (also likely in Ref. 8), hence the corrected $\Delta E_F$ values are 6.1 meV and 4.6 meV for Regions A and B respectively, whilst Region C remains essentially unchanged due to the small coherence length.

Figure 2(e) plots the measured ξ as a function of $E_F$. The shaded region shows the coherence length estimated within the Thomas-Fermi approximation $\xi = \sqrt{\frac{\pi}{2\alpha g}} \left(\frac{\hbar v_F}{E_F}\right)$ where the spin/valley degeneracy $g =$ 4, α is the fine structure constant and $v_F$ the Fermi velocity. The agreement is good within uncertainty but closer to the lower bound, implying a larger α (i.e. more strongly interacting system) consistent with previous studies.[19] Assuming α = 0.174, we can calculate an impurity density and mobility for Regions A, B and C using the Thomas-Fermi (TF) and Random Phase (RPA) approximations (see Section S6 in Supplementary Information for full calculation and results). For Region C we infer $n_{imp}$ of $3.6 \times 10^{18}$ cm$^{-3}$ from RPA, in good agreement with the doping measured by Hall effect ($4.35 \times 10^{18}$ cm$^{-3}$), demonstrating that the dopants are charged impurities. However, the experimental mobility (3540 cm$^2$V$^{-1}$s$^{-1}$) is significantly lower than expected (19,000 cm$^2$V$^{-1}$s$^{-1}$) for charged impurity scattering alone, indicating that other sources of disorder (e.g. point defects, grain boundaries) are important[19] and suggests that ultra-high mobility thin-film Na$_3$Bi could be achieved provided these other sources of disorder can be eliminated.

We now turn to the resonance feature (labelled D) observed in the STS spectra in Fig. 1(d). Figure 3(a) plots the peak differential conductance of the resonant feature D, with the red markers indicating the position of defect sites observed in the topography of Region A [Fig. 1(b)]. The high degree of correlation indicates that the defect sites are the origin of the resonance feature (see Fig. S3 in SI for further analysis). These defects correspond to Na surface vacancies (inset of Fig. 1(a)) in position Na(2) of the crystal structure of Na$_3$Bi shown in Fig. 3(b). To better understand the origin of the resonance feature, DFT calculations including spin-orbit coupling were performed (see Methods). Figure 3(c) compares an experimental STS for Na$_3$Bi on Si(111) (green curve) with the calculated density of states $D(E)$ for the following Na$_3$Bi structures with an Na(2) vacancy: Bulk Na$_3$Bi with one Na(2) vacancy per 8 unit cells (black curve), a Na$_3$Bi slab (2x2x2 unit cells) containing an Na(2) vacancy in the interior of the slab (blue curve), and a Na$_3$Bi slab (2x2x3 unit cells) containing an Na(2) vacancy at the surface (red curve). The associated bandstructures for each of the structures are shown in Fig. 3(d). In all cases the Na(2) vacancy accepts one electron. Additionally we find that indeed Na(2) vacancies at the surface produce a peak in $D(E)$ at an energy close to the experiment (red curve), due to the formation of a "Mexican hat" shaped valence band edge (see Fig. 3(d)). This structure and associated D(E) peak are not present for Na(2) vacancies in the bulk or in the interior of confined slabs (see additional discussion in Supplementary Information). Since the enhanced $D(E)$ at the surface results from a bandstructure effect, we expect it is not localised at defect sites but rather varies on a length scale set by the Fermi wavelength, in agreement with observations. The Na(2) surface vacancy concentration thus provides a unique knob to tune the surface density of states of this Dirac material, which could be used e.g. to tune electron-electron interactions, or tune the coupling to magnetic impurities.

Qualitatively different behavior was observed for occasional defect clusters. Figure 4(a) where the STM topography of a 60 nm x 60 nm region of Na$_3$Bi shows a large concentration of singular defects sites, as well as one defect which appears in topography to consist of a multi-vacancy cluster. The spectroscopic signature of such defects is very different from the quasi-bound resonant state of the singular defects [see Figure 3], as shown in the fixed-bias d$I$/d$V$ maps (b)-(d) taken at -196 meV, -216 meV and -236 meV respectively. Due to long-ranged electrostatic interaction of the defect with the scanning tip, a ring-like structure emerges that increases in spatial extent as the tip-sample potential becomes more negative. This phenomenon has also been observed for defects in graphene[21], and dopants in semiconductor systems[22] as a tip-induced 'ionization charging ring', where the sudden increase in charge of a bound defect state at a particular tip potential affects the screening cloud in the substrate at the tip position. In this case the defect state is truly localized, in contrast to the quasi-bound state observed for single Na vacancies, and offers an opportunity to study a quantum dot connected to a TDS reservoir.

We have demonstrated using scanning tunneling microscopy and spectroscopy the existence of charge puddles in the topological Dirac semimetal, $Na_3Bi$. The ultra-low Dirac-point energy fluctuations, which occur over length scales of approximately 5-15 nm, are of the order of 4-6 meV = 40-70 K well below room temperature and comparable to the highest quality graphene on h-BN. The ultra-low potential fluctuations in this 3D Dirac system will enable the exploration of novel physics associated with the Dirac point.[1-5] In addition, we observed for the first time defect-associated quasi-bound and bound states in a 3D TDS, which open the possibility to tune electron-electron interactions, or tune the coupling to magnetic impurities by varying the sodium surface vacancy concentration.

**Experimental Methods:**
The 20 nm $Na_3Bi$ thin film was grown in a ultra-high vacuum (UHV) ($10^{-10}$ Torr) molecular beam epitaxy (MBE) chamber and then transferred immediately after the growth to the interconnected Createc LT-STM operating in UHV ($10^{-11}$ Torr) for STM/STS measurements at 5 K. For $Na_3Bi$ film growth effusion cells were used to simultaneously evaporate elemental Bi (99.999%, Alfa Aesar) in an overflux of Na (99.95%, Sigma Aldrich) with a Bi:Na flux ratio not less than 1:10, calibrated by quartz microbalance. The Bi rate used was ~0.03 Å/s, and Na was ~0.7 Å/s. The pressure during growth was less than $3 \times 10^{-9}$ Torr.

*Growth on Si(111)* - To prepare an atomically flat substrate, a Si(111) wafer was flash annealed in order to achieve 7 x 7 surface reconstruction, confirmed using STM. During the growth, the substrate temperature was 330 $^0$C for successful crystallization. At the end of the growth the sample was left at 330 $^0$C for 10 min in a Na overflux to improve the film quality before cooling to room temperature.

*Growth on Sapphire* - The sapphire substrate was annealed in atmosphere at 1350$^o$C and then pure oxygen atmosphere at 1050$^o$C to achieve an atomically flat surface. Ti/Au (5/50nm) contacts were deposited on the corners of the substrate, and wirebonded to a contact busbar on the sample plate to allow for *in-situ* transport measurements in a 1T perpendicular magnetic field at 5K. The sapphire was annealed in UHV to 400$^o$C for 1 hour to remove atmospheric species. $Na_3Bi$ films were then grown using a two-step growth method as reported previously in [Ref. 19]. The $Na_3Bi$ film used in this study had a final growth temperature was 330$^o$C.

A PtIr STM tip was prepared and calibrated using an Au(111) single crystal and the Shockley surface state before all measurements. STM differential conductance ($dI/dV$) was measured using a 5 mV AC excitation voltage (673 Hz) that was added to the tunnelling bias. Differential conductance measurements were made under open feedback conditions with the tip in a fixed position above the surface. Data was prepared and analysed using MatLab and WSxM software.[24]

**Density functional theory methods:**
First-principles calculations based on density-functional theory are used to obtain electronic bands and density of states of bulk, bilayer and trilayer $Na_3Bi$, with and without Na(2) vacancies. The first-principles approach is based on Kohn-Sham density functional theory (KS-DFT),[25] as implemented in the Quantum ESPRESSO code.[26] The exchange correlation energy is described by the generalized gradient approximation (GGA) using the PBE functional.[27] Interactions between valence and core electrons are described by Troullier-Martins pseudopotentials.[28] The Kohn-Sham orbitals were expanded in a plane-wave basis with a cutoff energy of 50 Ry, and for the charge density, a cutoff of 200 Ry was used. The Brillouin-zone (BZ) was sampled using Γ-centered grids, following the scheme proposed by Monkhorst-Pack.[29] For convergence of charge density, an 8x8x4 grid was used for bulk while a 6x6x1 grid was used for bilayer and trilayer. For the density of states, a finer grid of 32x32x16 (26x26x1) is employed for bulk (bilayer and trilayer). The calculation of the spin-orbit splitting was performed using non-colinear calculations with fully relativistic pseudopotentials. For the bilayer and trilayer models, we used periodic boundary conditions along the three dimensions, with vacuum regions of 10 Å between adjacent images in the direction perpendicular to the layers. Convergence tests with greater vacuum spacing, guarantee that this size is enough to avoid spurious interactions between neighbouring images.

**Acknowledgements**

J. L. C., M. T. E., J. H and M. S. F. are supported by M. S. F.'s ARC Laureate Fellowship (FL120100038). M. T. E. is supported by ARC DECRA fellowship DE160101157. S. A., I. Y. and J. N.

B. R. are supported under the National Research Foundation of Singapore's fellowship program (NRF-NRFF2012-01). L. C. G. acknowledges A. H. Castro Neto and is supported by the National Research Foundation-Prime Minister's Office Singapore, under its Medium-Sized Centre funding. The first-principles calculations were carried out on the CA2DM high-performance computing facilities.

**Author Contributions**

M. T. E. and J. L. C. contributed equally to this work. M. T. E., J. H., and M. S. F. devised the experiments. M. T. E. and J. L. C. performed the MBE growth and STM/STS measurements. I. Y. and S. A. assisted with the theoretical understanding and interpretation of the charge puddling. L. C. G and J. N. B. R. performed the DFT calculations. J. L. C., M. T. E., and M. S. F. analyzed the data and composed the manuscript.

**Additional Information**

The authors declare no competing financial interests. Supplementary information can be found.

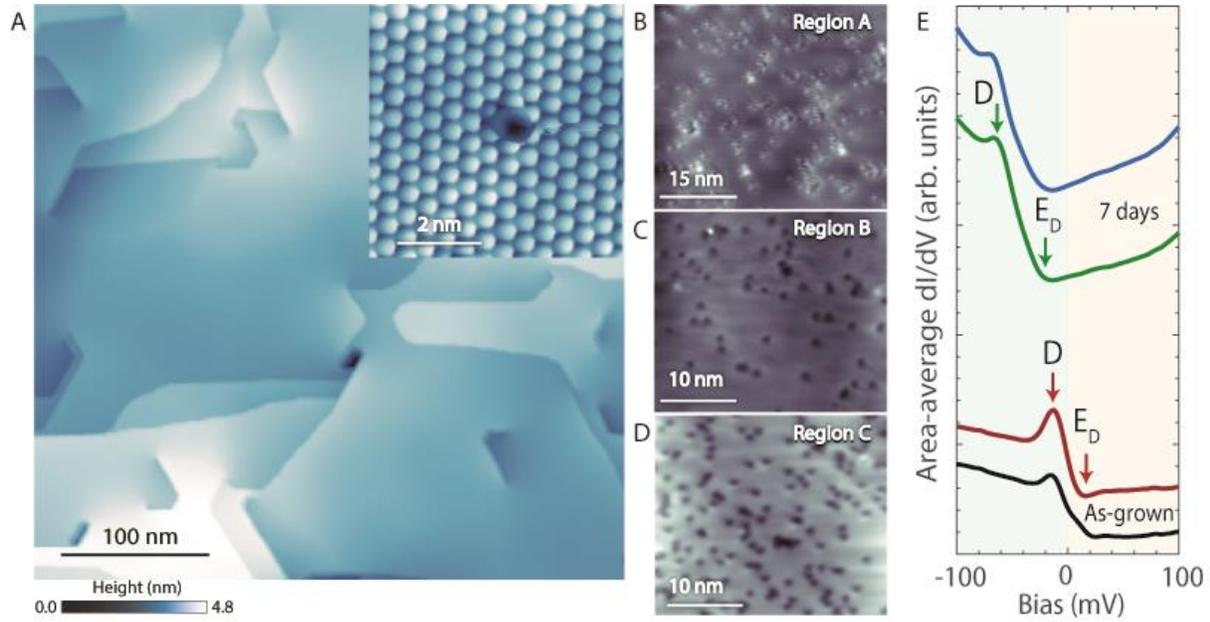

**Figure 1. Large area morphology and spectroscopy of 20 nm Na₃Bi on Si(111).** (A) Large area (400nm x 380 nm) topographic STM image (bias voltage $V$ = -3 V and tunnel current $I$ = 50 pA). Inset: Atomic resolution STM image with lattice constant 5.45 Å (taken on separate 20 nm Na₃Bi film) showing an individual Na vacancy at the surface. (B) STM topography ($V$ = 300 mV and $I$ = 250 pA) on a 45 nm x 45 nm region of Na₃Bi. (C) STM topography *($V$ = -550 mV and $I$ = 200 pA)* on a 30 nm x 30 nm region of Na₃Bi. (D) STM topography ($V$ = -50 mV and $I$ = 100 pA) on a 30 nm x 30 nm region of Na₃Bi on sapphire (α-Al₂O₃(0001). (E) Area-averaged STS spectra (vertically offset for clarity) corresponding to four different regions of the sample. Black spectra corresponding to the region in (B) and red spectra (topography not shown) were taken immediately after growth, whilst the blue spectra corresponding to the region in (C) and green spectra (topography not shown) were taken one week after growth. Feature $E_D$ reflects the Dirac point, which corresponds to the minimum in the LDOS, and D represents the resonance feature (discussed in text).

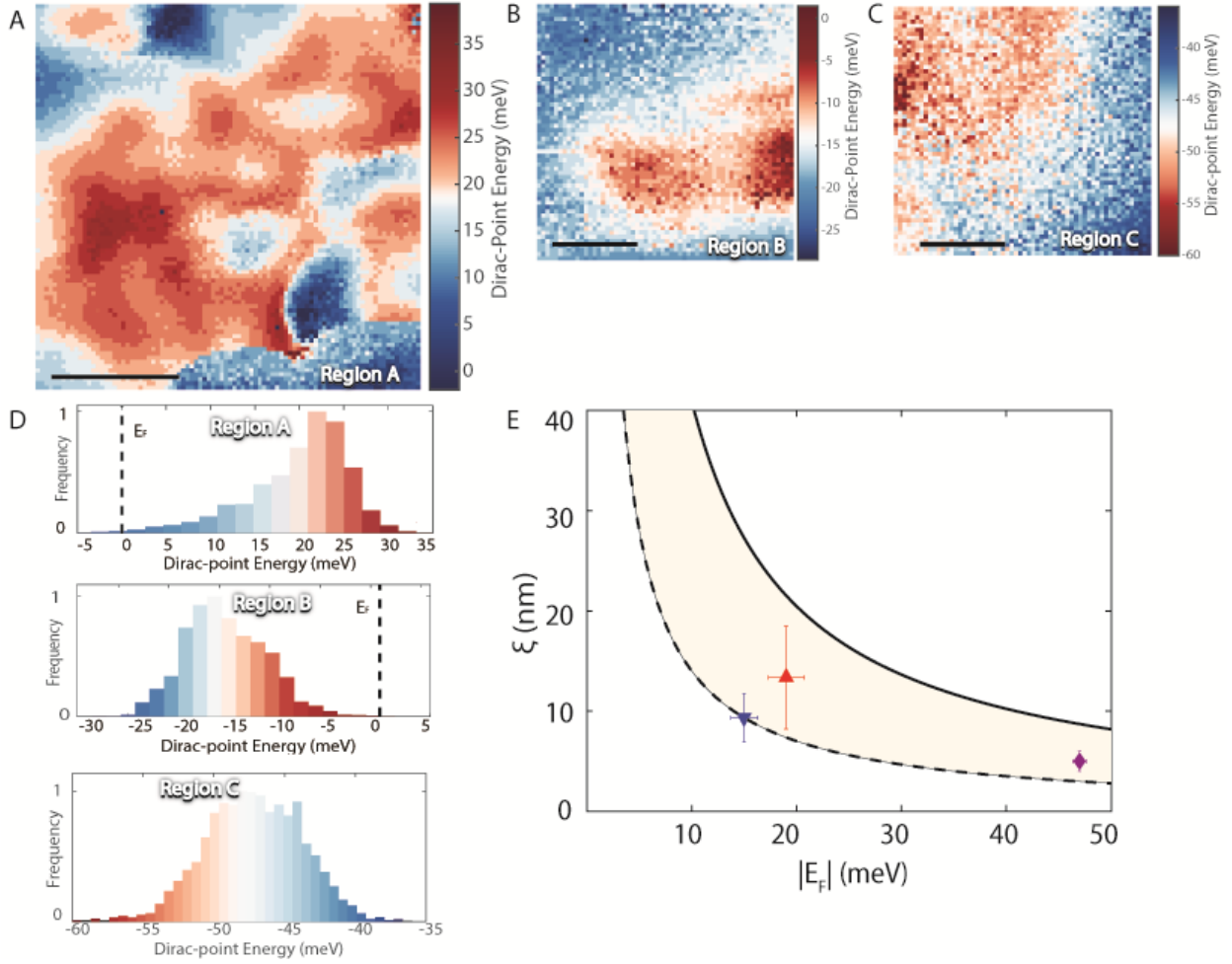

**Figure 2. Charge puddling profiles of *p*-type and *n*-type Na₃Bi.** (A) Dirac point energy map of 45 nm x 45 nm (90 pixels x 90 pixels) region of Na$_3$Bi ($V$ = -250 mV and $I$ = 250 pA), corresponding to Region A represented in Figure 1(B). Scale bar is 15nm. (B) Dirac-point energy map of 30 nm x 30 nm (60 pixels x 60 pixels) region of Na$_3$Bi corresponding to Region B of Figure 1(C) ($V$ = -150 mV and $I$ = 200 pA). Scale bar is 10 nm. (C) Dirac-point energy map of 30 nm x 30 nm (60 pixels x 60 pixels) region of Na$_3$Bi grown on α-Al$_2$O$_3$(0001) (labelled Region C) ($V$ = -150 mV and $I$ = 200 pA). Scale bar is 10 nm. (D) Upper, middle and lower panels representing histograms of the Dirac point energy maps in (A)-(C) respectively. The histograms are colour-coded to reflect the intensity scale in the corresponding Dirac-point energy map. (E) Plot of spatial coherence length as a function of Fermi energy, comparing the experimental data for Region A (red triangle), Region B (blue triangle) and Region C (purple diamond) to theoretical predictions (shaded region) where the upper bound (solid line) is defined using $v_F$ = 2.4x10$^5$ ms$^{-1}$ and α = 0.069, whilst the lower bound (dashed line) uses $v_F$ = 1.4x10$^5$ ms$^{-1}$ and α = 0.174[19].

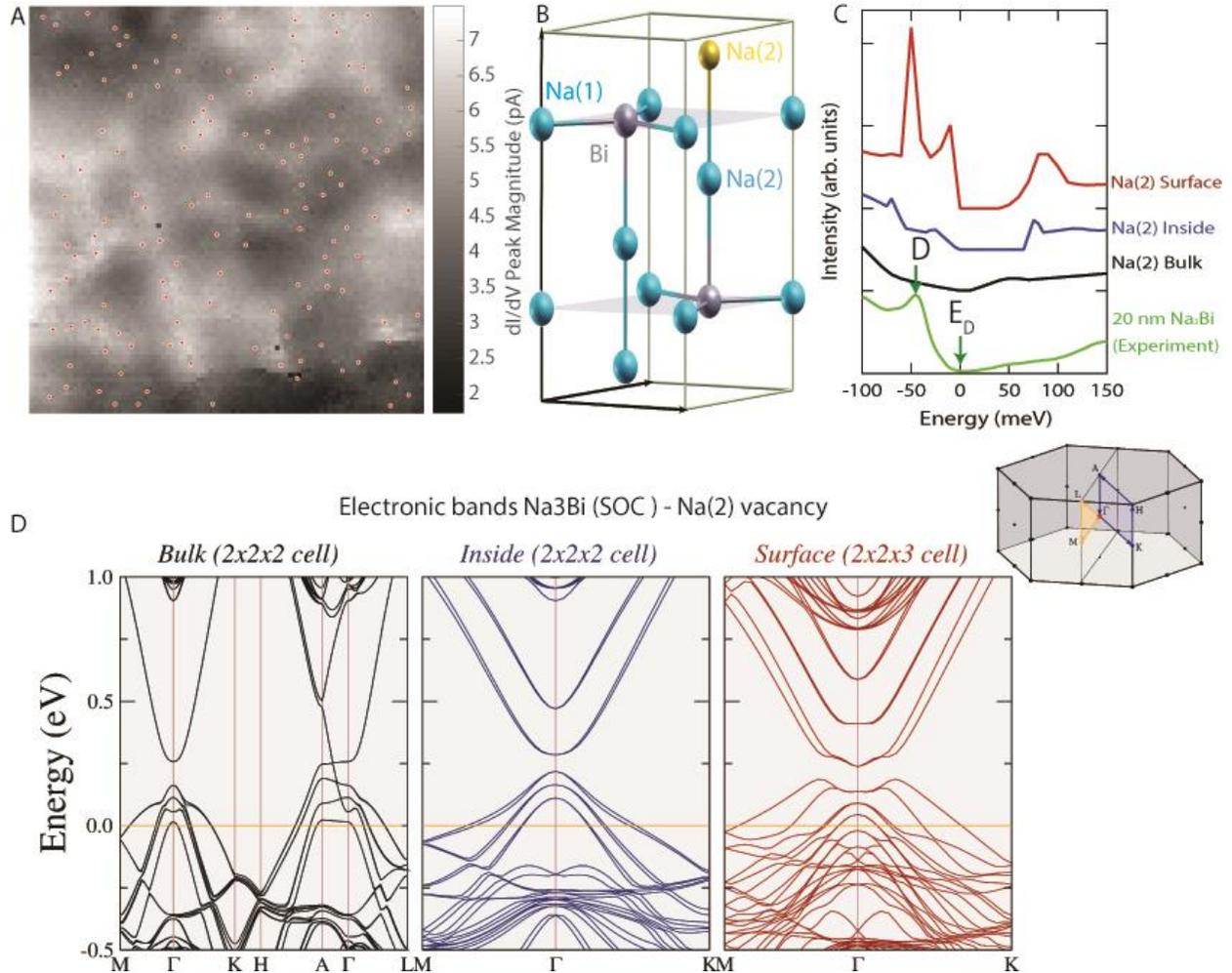

**Figure 3. Determining the bound state defect resonance.** (A) Map of the d*I*/d*V* magnitude at the defect resonance energy, where defects are shown as red circles. (B) Crystal structure of Na₃Bi, with the surface terminated Na labelled Na(2)  (gold), with the remaining Na atoms in blue with the Na bonded to Bi in the hexagonal lattice labelled Na(1) and the Bi atoms in grey. (C) Comparison between DFT calculations of the DOS for Na₃Bi with a Na(2) vacancy at the surface of a 2x2x3 cell (red curve), a Na(2) vacancy inside a 2x2x2 cell (blue) and bulk Na(2) vacancy (black curve) and the experimental STS curve for a 20 nm Na₃Bi film (green curve). Energy scales of all spectra have been corrected so that 0 eV reflects the Dirac point. A vertical offset has been applied for clarity. (d) Accompanying electronic bandstructures for bulk Na₃Bi with a Na(2) vacancy (left panel), Na(2) vacancy inside a 2x2x2 cell (middle panel) and Na(2) vacancy at the surface of a 2x2x3 cell (right panel). Inset: Brillouin zone of bulk Na₃Bi and the projected (001) surface.

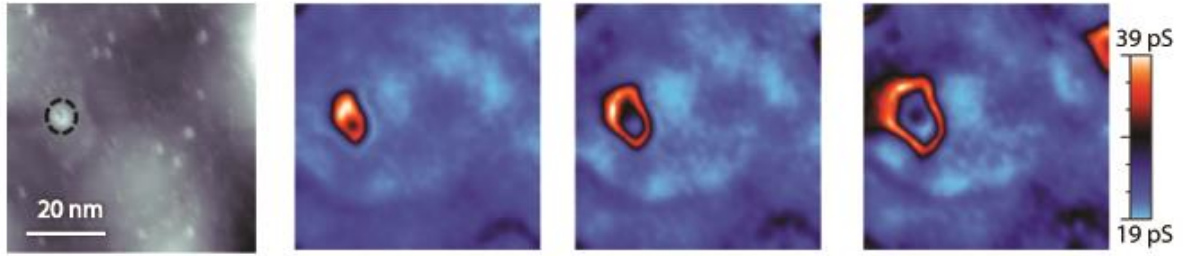

**Figure 4. Tip-induced ionization rings around large defects.** (A) STM topography (V = -250 mV and I = 250pA) on a 60 nm x 60 nm region of Na$_3$Bi. Fixed bias d$I$/d$V$ maps taken at (B) -196 mV, (C) -216 mV and (D) -236 mV over the same region as (A) showing a ring like feature centred around a large vacancy site highlighted by the dashed circle in (A).

**Supplementary Information**

**Spatial Charge Inhomogeneity and Defect States in Topological Dirac Semimetal Thin Films**

Mark T. Edmonds[1,2†*], James L. Collins[1,2†], Jack Hellerstedt[1,2], Indra Yudhistira[3], Lídia C. Gomes[3], João N. B. Rodrigues[3], Shaffique Adam[3,4], Michael S. Fuhrer[1,2*]

1.  School of Physics and Astronomy, Monash University, Clayton VIC 3800, Australia
2.  Monash Centre for Atomically Thin Materials, Monash University, Clayton VIC 3800 Australia
3.  Department of Physics and Centre for Advanced 2D Materials, National University of Singapore, 117551, Singapore
4.  Yale-NUS College, 6 College Avenue East, 138614, Singapore

† These authors contributed equally to this work

* Corresponding author: mark.edmonds@monash.edu and michael.fuhrer@monash.edu

**TABLE OF CONTENTS:**

**S1. Determining the Dirac point position from the dI/dV spectra**

For a Dirac semimetal such as graphene, or in the present case of 3D Dirac semimetal $Na_3Bi$, the density of states (DOS) should be at a minimum at the Dirac-point Energy ($E_D$). This manifests in the scanning tunneling spectra as a minima in the differential conductance (dI/dV) as a function of sample bias, which is proportional to the local density of states. The simplest approach to determining the minima in the dI/dV (i.e. Dirac point) is fitting the spectra to a low order polynomial. However, in the case of $Na_3Bi$ the presence of a defect-associated bound energy resonance ~33meV below $E_D$ means standard polynomial fitting does not always yield accurate determination of the Dirac point as shown in Fig. S1. To counter this problem a non-parametric local regression procedure in MATLAB was employed to produce fits of the dI/dV spectra to then identify and track critical points of the DOS (i.e. $E_D$ and also the defect resonance) with reduced systematic error. This procedure then allowed us to construct puddling maps from either (1) the change in the minima directly, or (2) from the change in the defect resonance position D (which also reflect changes in $E_D$) and then correcting for the constant energy separation of $E_D - D = 35$ meV.

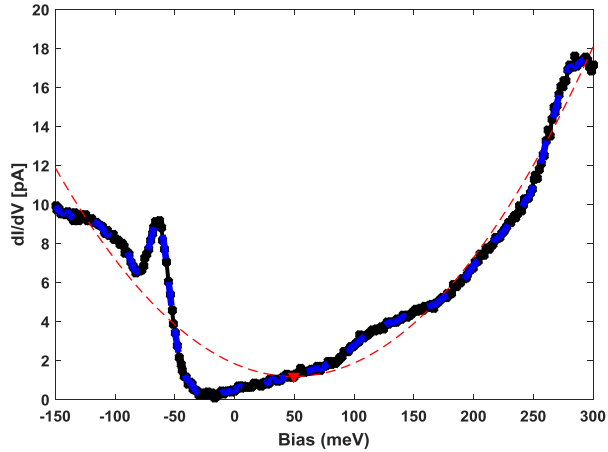

**Figure S1. Determining the Dirac point position from the dI/dV spectra.** A single dI/dV spectra (black curve) taken at an arbitrary point on an atomically flat region of $Na_3Bi$ on Region B (Fig. 1(c) main manuscript). This spectra is fit to a low-order polynomial (red curve) and a non-parametric local regression (blue curve). The large discrepancy between the minima position in the dI/dV spectra and the minima obtained from the low-order polynomial highlights the need for a detailed fitting procedure to ensure accurate determination of Dirac point position.

## S2. Demonstrating that charge puddling is tied to Na(2) vacancies

In order to demonstrate that the defects attributed to Na vacancies of the lattice in $Na_3Bi$ are correlated with the fluctuations in the energy potential we compare the STM topography and charge puddle map of Region A in Fig. S2(a) and (b) respectively. Black circles are used to highlight the defect sites in the charge puddles map in Fig. S2(b). In Fig. S2(c) histograms representing the entire Region A (orange stairs) and only the black marker locations (blue bars) are plotted. A shift to higher energy of ~4 meV is observed. This is consistent with our DFT calculations which indicate that the Na(2) vacancy accepts one electron (see Fig. S5). We conclude that Na(2) vacancies acts as negatively charged impurities, increasing the local hole doping.

In order to determine the expected doping from the sodium vacancies we compare the number of defects observed in the topography with the predicted impurity density calculated using the random phase approximation. In Fig. S2(a) ~133 sodium vacancies are located within the 45 nm x 45 nm scan region, equating to $6.6x10^{12}$ sodium vacancies per $cm^2$, giving $3.3x10^{18}$ $cm^{-3}$ for 20 nm of $Na_3Bi$ (a similar defect concentration is found for Region B). This is similar to the predicted impurity density of 2.9-4.5 $x10^{18}$ $cm^{-3}$ calculated for Region A in Table S1. This demonstrates that sodium vacancies are a major source of the impurity density and disorder in pristine $Na_3Bi$ films grown on Si(111). However, it should be noted for the *n*-type doped $Na_3Bi$ there are clearly other sources of doping than sodium vacancies which are acceptors. At present we cannot determine the exact source of the large *n*-type doping, or whether those dopants lie at the substrate interface or distributed throughout the bulk of the $Na_3Bi$ film.

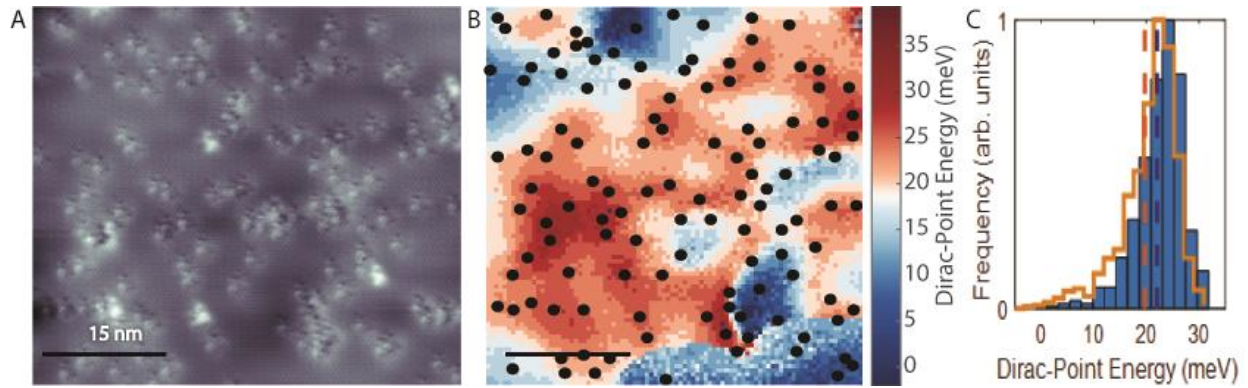

**Figure S2**. **STM topography and charge puddling map of *p*-type $Na_3Bi$ on Region A.** (A) STM topography corresponding to Region A (Fig. 1(B)) of the main manuscript. (B) Charge puddling map corresponding to Fig. 2(A) of the main manuscript, with black circles representing the defect sites. (C) Two histograms of subsets of the Dirac-point energy in Region A, which have been normalized for clarity. The orange stairs represent the Dirac-point energy over all of Region A (with the mean of 20 meV, marked as a dashed orange line) and the blue bars represent the Dirac-point energy measured only at the black marker locations (with the mean of 24 meV marked as a blue dashed line).

## S3. Correlation between resonance state in dI/dV spectra with lattice defects

To further demonstrate the relationship between the magnitude of the bound resonant state and lattice defects we examine the defect resonance and lattice defects of Region A in Fig. S3. In Fig. S3(a) the dI/dV resonance intensity is plotted for Region A, along with markers highlighting the defect positions. In Fig. S3(b) we plot the dI/dV resonance histograms of all of Region A (orange stairs) and only the defect sites marked by green circles (blue bars). We omit the defects highlighted in red due to those regions showing faint signatures of charging ring behaviour, which affects the resonance intensity. The resonance differential conductance magnitude over all of Region A has a mean value at 4.9pA and standard deviation of 0.90pA, whilst at only the green defect locations there is a mean value of 5.4pA with standard deviation 0.66pA. This shift to higher dI/dV intensity at the defect sites highlights the direct correlation between the resonance state and defect location.

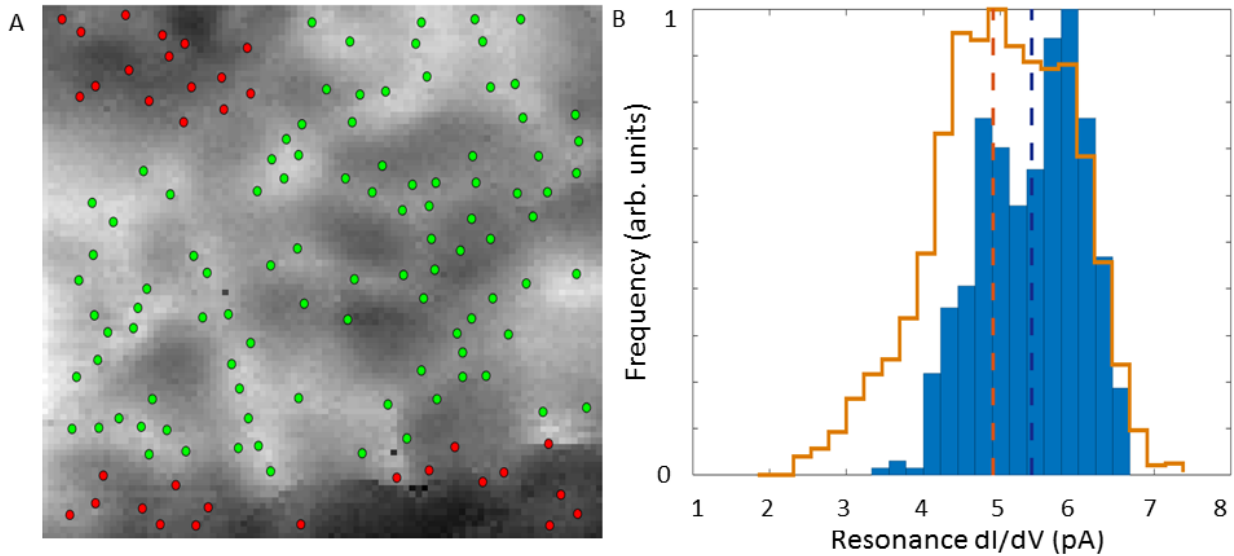

**Figure S3**. **Spatial dependence of defect resonance.** (A) Map of the dI/dV magnitude at the defect resonance energy corresponding to Fig. 3(A) in the main manuscript. Markers indicate the location of lattice defects, with the defects assigned red and green marker, based on whether their location shows (red) and does not show (green) faint signatures of charging-ring behaviour. (B) Two histograms of subsets of the resonance magnitude in Region A, which have been normalized for clarity. The orange stairs represent the resonance differential conductance magnitude over all of Region A (with the mean of 4.9pA marked as a dashed orange line) and the blue bars represent the resonance differential conductance measured only at the green marker locations (with the mean of 5.4pA marked as a blue dashed line).

**S4. Calculation of correlation length, impurity density and mobility**

The coefficients of an autocorrelation matrix $C(x_1, x_2)$ for charge puddle fluctuations in $E_D$, is predicted to exhibit exponential spatial dependence in two-dimensions. With the assumption that there should be no necessary anisotropy of the measurement, it is convenient to use a radially-averaged $C(R)$. The average coherence length of $C(R)$ can then be taken as the puddle length scale $\xi$; this is the distance at which $C(R)$ = $C(0)(\frac{1}{e} - 1)$, where $e$ is Euler's number.

As the experimental length scale L is not much larger than the observed fluctuation length scale, it is suspected that undersampling is responsible for producing anti-correlation coefficients in $C(R)$ at length scales greater than $L^2/\pi$. We attempt to account for this in two ways:

(1) We perform an exponential regression over the range L, which does not account for undersampling-related uncertainty in $C(R)$.

(2) We perform linear regression of $\log C(R)$, weighted by an estimate for error $\delta C(R)$. On a length scale R, there are $L^2 - R^2$ pairs of measurements which are used for C(R), however the number of statistically different measurements is of order $1 + (L^2 - R^2)/\xi^2$. Taking $\delta C(R)$ then as the inverse standard deviation of measurements, $\delta C(R) \cong \left(1 + \frac{L^2 - R^2}{\xi^2}\right)^{-2}$.

By performing least-squares fitting of $\log C(R)$ over the range of positive $C(R)$, we achieve an estimate of $\xi$ from goodness of fit analysis after $\delta C(R)$ for a given $\xi$ propagates. By taking a lower-bound ($\chi^2 = 1$ best fit) and upper-bound ($1\sigma$ overlap of fit to model with error), the mean of these two values is then our estimate for $\xi$.

Additionally, as charge puddles are expected to follow a Gaussian energy distribution, we place an estimate of $\delta|E_F|$ as the standard error, where mean $\sigma_{mean} = \sigma/\sqrt{N}$, and further assume that the number of measurements is $N = L^2/\xi^2$.

## S5. Theory discussion on correlation length, impurity density and mobility

To theoretically compute the correlation functions in Na$_3$Bi, we make two assumptions. First, the impurity potential comes from a random three-dimensional (3D) distribution of charged impurities with density $n_{\text{imp}}$. Second, we assume that it is possible to find a global screening function $\varepsilon(q, n_{\text{eff}})$ that can describe the effects of these local screening variations. With these assumptions, we can write the autocorrelation function of the screened Coulomb potential as

$$C(r) = n_{\text{imp}} \int d^3 q \left| \frac{V(q)}{\varepsilon(q, n_{\text{eff}})} \right|^2 e^{iq \cdot r}$$

$$= 8 n_{\text{imp}} \left( \frac{e^2}{\kappa} \right)^2 \int_0^\infty dq \frac{q^2}{\left[ q^2 \varepsilon(q, n_{\text{eff}}) \right]^2} \frac{\sin(qr)}{qr}$$

$$= 8 n_{\text{imp}} \left( \frac{e^2}{\kappa} \right)^2 \int_0^\infty dq \frac{q^2}{\left( q^2 + q_s^2 \right)^2} \frac{\sin(qr)}{qr}$$

$$= \frac{4 n_{\text{imp}}}{k_{\text{eff}}} \left( \frac{e^2}{\kappa} \right)^2 I \left( \sqrt{\frac{g\alpha}{2\pi}}, r \right) \tag{1}$$

$$I(z, r) = \int_0^\infty d\eta \frac{\eta^2}{\left[ \eta^2 + z^2 \tilde{\Pi}(2k_{\text{eff}} \eta) \right]^2} \frac{\sin(2k_{\text{eff}} \eta r)}{2 k_{\text{eff}} \eta r}, \tag{2}$$

where $n_{\text{imp}}$ is impurity density, $r$ is the spatial distance between two points, and $\eta = q/(2k_{\text{eff}})$.
Here we have used long-range Coulomb potential

$$V(q) = \frac{4\pi e^2}{\kappa q^2}, \tag{3}$$

where $-e$ is the electron charge, $\kappa$ is the effective dielectric function of the material, and $q$ is the momentum transfer between incoming and outgoing momenta of the scattered electron.
The dielectric function is

$$\varepsilon(q, n_{\text{eff}}) = 1 + V(q) \Pi(q)$$

$$= 1 + V(q) \nu(n_{\text{eff}}) \tilde{\Pi}(q)$$

$$= 1 + \left[ \frac{q_s(n_{\text{eff}})}{q} \right]^2, \tag{4}$$

where $\Pi(q)$ is polarizability function, $\nu(n_{\text{eff}}) = \frac{g}{2\pi^2} \frac{k_{\text{eff}}^2}{\hbar v_F}$ is the density of states of Na$_3$Bi, and $\tilde{\Pi}(q) \equiv \Pi(q)/\nu(n_{\text{eff}})$. Introducing effective fine structure constant $\alpha = e^2/(\hbar v_F \kappa)$, we get the inverse screening length $q_s(n_{\text{eff}}) = \sqrt{2 g \alpha \tilde{\Pi}(q)/\pi} k_{\text{eff}}$. Here $g$ is degeneracy factor, $v_F$ is Fermi velocity and $k_{\text{eff}}$ is effective Fermi wavevector.

## 1.1 Thomas– Fermi

Thomas-Fermi polarizability is given by $\Pi_{\text{TF}}(q) = \nu(n_{\text{eff}})$ or $\tilde{\Pi}_{\text{TF}}(q) = 1$, hence $q_{\text{TF}} = \sqrt{2g\alpha/\pi}\,k_{\text{eff}}$. In this case $I_{TF}(z, r)$ can be evaluated analytically as

$$I_{TF}(z, r) = \int_0^\infty d\eta \, \frac{\eta^2}{\left(\eta^2 + z^2\right)^2} \frac{\sin(2k_{\text{eff}}\eta r)}{2k_{\text{eff}}\eta r}$$

$$= \frac{\pi}{4z} e^{-2k_{\text{eff}}zr}. \tag{5}$$

Hence the autocorrelation function can be calculated as

$$C_{\text{TF}}(r) = \frac{4n_{\text{imp}}}{k_{\text{eff}}} \left(\frac{e^2}{\kappa}\right)^2 I_{\text{TF}}\left(\frac{q_{\text{TF}}}{2k_{\text{eff}}}, r\right)$$

$$= \frac{2\pi n_{\text{imp}}}{q_{\text{TF}}(n_{\text{eff}})} \left(\frac{e^2}{\kappa}\right)^2 e^{-q_{\text{TF}}(n_{\text{eff}})r}. \tag{6}$$

We would like to remark that the autocorrelation function has an exponential spatial profile, in contrast with autocorrelation function in monolayer graphene, that has a gaussian profile.[1,2] We can then define Thomas-Fermi correlation length as nothing but Thomas-Fermi effective screening length

$$\xi_{\text{TF}}(n_{\text{eff}}) = 1/q_{\text{TF}}(n_{\text{eff}})$$

$$= \sqrt{\frac{\pi}{2g\alpha}} \frac{1}{k_{\text{eff}}}.$$

Since in our experiment the Fermi energy $E_F$ is larger than the fluctuation in energy $E_{\text{rms}}$, effective wavevector can very well be approximated by Fermi wavevector $k_{\text{eff}} \approx k_F$, leading to

$$\xi_{\text{TF}} \approx \sqrt{\frac{\pi}{2g\alpha}} \frac{1}{k_F}$$

$$= \sqrt{\frac{\pi}{2g\alpha}} \frac{\hbar v_F}{|E_F|}. \tag{7}$$

Here, we have used effective low energy dispersion relation of Na$_3$Bi, $E_F = sgn(n)\hbar v_F k_F$ in the last line. We plotted the upper bound and lower bound of the Thomas-Fermi correlation length plotted in figure 2(d) of the main paper using the formula above with $v_F = 2.43 \times 10^5$ m/s, $\kappa_{\text{eff}} = 130$ and $v_F = 1.4 \times 10^5$ m/s, $\kappa_{\text{eff}} = 90$,[3] respectively. We have used $g = 4$ to take spin and valley degeneracy into account.

## 1.2 Random Phase Approximation

The ratio of the random phase approximation (RPA) polarization function and the density of states $\tilde{\Pi}_{\text{RPA}}[x = q/(2k_F)]$ is given by the sum of two components: a vacuum part $\tilde{\Pi}_V(x)$ and a finite density part $\tilde{\Pi}_M(x)$ [4]

$$\tilde{\Pi}_V = \frac{2}{3}\left[1 + \frac{1}{4x}(1 - 3x^2)\log\left|\frac{1+x}{1-x}\right| - \frac{x^2}{2}\log\left|\frac{1-x^2}{x^2}\right|\right]$$

$$\tilde{\Pi}_{\mathrm{M}} = \frac{2x^2}{3} \log \left| \frac{\delta}{x} \right|. \tag{8}$$

Here $\delta = \Delta/(2k_F)$, where $\Delta$ is ultraviolet cutoff.

Analytical evaluation of RPA autocorrelation function is not possible. However, the numerics of RPA autocorrelation function is found to also have exponential decay with spatial distance $r$ just like Thomas–Fermi autocorrelation function. Hence, we can obtain the RPA correlation length $\xi_{\mathrm{RPA}}$ by fitting the numerics of the ratio of RPA autocorrelation function at finite $r$ and at $r = 0$ to exponential function

$$\frac{C_{\mathrm{RPA}}(r)}{C_{\mathrm{RPA}}(r=0)} \equiv \exp\left(-\frac{r}{\xi_{\mathrm{RPA}}}\right). \tag{9}$$

We have found that RPA correlation length doesn't differ much from Thomas–Fermi correlation length.

## 2 Impurity density

The fluctuation in the screened, charged impurity-induced potential $E_{\mathrm{rms}}$ is given by

$$E_{\mathrm{rms}}^2 = \delta V^2$$
$$= C(r=0)$$
$$= n_{\mathrm{imp}} \int \frac{d^3 q}{(2\pi)^3} \left| \frac{V(q)}{\varepsilon(q, n_{\mathrm{eff}})} \right|^2$$
$$= 8 n_{\mathrm{imp}} \left( \frac{e^2}{\kappa} \right)^2 \int_0^\infty dq \frac{q^2}{\left(q^2 + q_s^2\right)^2}$$
$$= 4\alpha^2 J\left( \sqrt{\frac{g\alpha}{2\pi}}, k_{\mathrm{eff}} \right) \frac{(\hbar v_F)^2 n_{\mathrm{imp}}}{k_{\mathrm{eff}}} \tag{10}$$

$$J(z,k) = \int_0^\infty d\eta \frac{\eta^2}{\left[\eta^2 + z^2 \tilde{\Pi}(2k\eta)\right]^2}. \tag{11}$$

Since in our experiment the Fermi energy $E_F$ is larger than the fluctuation in energy $E_{\mathrm{rms}}$, effective wavevector can very well be approximated by Fermi wavevector $k_{\mathrm{eff}} \approx k_F$, leading to

$$E_{\mathrm{rms}}^2 = 4\alpha^2 J\left( \sqrt{\frac{g\alpha}{2\pi}}, k_F \right) \frac{(\hbar v_F)^2 n_{\mathrm{imp}}}{k_F}$$
$$= 4\alpha^2 J\left( \sqrt{\frac{g\alpha}{2\pi}}, k_F \right) \frac{(\hbar v_F)^3 n_{\mathrm{imp}}}{|E_F|}, \tag{12}$$

where we have used effective low energy dispersion relation of $Na_3Bi$, $E_F = sgn(n)\hbar v_F k_F$ in the second line. Inverting this relation, we obtain charged impurity density

$$n_{\mathrm{imp}} = \frac{1}{4\alpha^2 J\left( \sqrt{\frac{g\alpha}{2\pi}}, k_F \right)} \frac{E_{\mathrm{rms}}^2 |E_F|}{(\hbar v_F)^3}. \tag{13}$$

## 2.1 Thomas– Fermi

Thomas– Fermi dimensionless polarizability is given by $\tilde{\Pi}_{\text{TF}}(q) = 1$, hence

$$J_{\text{TF}}(z,k) = \int_0^\infty d\eta \, \frac{\eta^2}{\left(\eta^2 + z^2\right)^2} = \frac{\pi}{4z}. \tag{14}$$

Therefore, charged impurity density for $E_F \gg E_{\text{rms}}$ is given by

$$n_{\text{imp}}^{\text{TF}} = \sqrt{\frac{g}{2(\pi\alpha)^3}} \frac{|E_F| E_{\text{rms}}^2}{(\hbar v_F)^3}. \tag{15}$$

## 2.2 Random Phase Approximation

Within RPA, the charged impurity density for $E_F \gg E_{\text{rms}}$ is given by

$$n_{\text{imp}} = \frac{1}{4\alpha^2 J_{\text{RPA}}\left(\sqrt{\frac{g\alpha}{2\pi}}, k_F\right)} \frac{E_{\text{rms}}^2 |E_F|}{(\hbar v_F)^3} \tag{16}$$

$$J_{\text{RPA}}(z,k) = \int_0^{\Delta/(2k)} d\eta \, \frac{\eta^2}{\left[\eta^2 + z^2 \tilde{\Pi}_{\text{RPA}}(2k\eta)\right]^2}, \tag{17}$$

where $J_{\text{RPA}}$ need to be evaluated numerically. Both Thomas– Fermi and RPA upper and lower bound for charged impurity density for region A, B, and C are computed in table S1.

Table S1: Charged impurity density for $E_F \gg E_{\text{rms}}$ calculated using Thomas– Fermi and RPA for both region A, B, and C.

|  | Region A | | Region B | | Region C | |
|---|---|---|---|---|---|---|
|  | $\alpha = 0.069$ | $\alpha = 0.174$ | $\alpha = 0.069$ | $\alpha = 0.174$ | $\alpha = 0.069$ | $\alpha = 0.174$ |
| $n_{\text{imp}}^{\text{TF}}\left(10^{18}\text{cm}^{-3}\right)$ | 2.5 | 3.3 | 1.1 | 1.5 | 2.0 | 2.6 |
| $n_{\text{imp}}^{\text{RPA}}\left(10^{18}\text{cm}^{-3}\right)$ | 2.9 | 4.5 | 1.3 | 2.0 | 2.3 | 3.6 |

Here, we have used $g = 4$ to take into account valley and spin degeneracy and ultraviolet cutoff $\Delta/(2k_F) = 100$ for RPA.

## 3 Mobility

The conductivity of Na$_3$Bi is given by

$$\sigma = e^2 \frac{v_F^2 \tau(E_F)}{3} \nu(E_F), \tag{18}$$

where $\nu(E_F) = \dfrac{g}{2\pi^2}\dfrac{k_F^2}{\hbar v_F}$ is density of states at Fermi level and $\tau$ is transport scattering time.

The Boltzmann transport scattering time for Coulomb impurities within Born approximation is given by

$$\frac{\hbar}{\tau(E_F)} = 2\pi n_{imp}\int\frac{d^3\mathbf{k}'}{(2\pi)^3}\left|\frac{V(q)}{\varepsilon(q)}\right|^2\frac{1-\cos^2\theta}{2}\delta(E_F - E_{\mathbf{k}'})$$

$$= \frac{n_{imp}}{2\pi}\frac{k_F^2}{\hbar v_F}\int_0^\pi d\theta \left|\frac{V\!\left(2k_F\sin\dfrac{\theta}{2}\right)}{\varepsilon\!\left(2k_F\sin\dfrac{\theta}{2}\right)}\right|^2 \sin\theta\frac{1-\cos^2\theta}{2}$$

$$= \frac{\pi}{2}\frac{n_{imp}}{\hbar v_F k_F^2}\left(\frac{e^2}{\kappa}\right)^2\int_0^\pi d\theta\frac{\sin^2\dfrac{\theta}{2}}{\left[\sin^2\dfrac{\theta}{2} + \dfrac{g\alpha}{2\pi}\tilde{\Pi}\!\left(2k_F\sin\dfrac{\theta}{2}\right)\right]^2}\sin\theta\frac{1-\cos^2\theta}{2}$$

$$= 4\pi\frac{n_{imp}\hbar v_F\alpha^2}{k_F^2}H\!\left(\sqrt{\frac{g\alpha}{2\pi}}\right) \tag{19}$$

$$H(z) = \int_0^1 d\eta\frac{\eta^3(1-\eta^2)}{\left[\eta^2 + z^2\tilde{\Pi}(2k_F\eta)\right]^2}. \tag{20}$$

where $n_{imp}$ is charged impurity density, $q = |\mathbf{k} - \mathbf{k}'|$ is the momentum transfer between incoming plane-waves with wavevector $\mathbf{k}$ and outgoing wavefunctions with wavevector $\mathbf{k}'$ and $\eta = q/(2k_F)$.

Hence, the conductivity is given by

$$\sigma = \frac{e^2}{h}\frac{g}{12\pi^2\alpha^2}\frac{k_F^4}{n_{imp}}\frac{1}{H\!\left(\sqrt{\dfrac{g\alpha}{2\pi}}\right)}. \tag{21}$$

Therefore, mobility is

$$\mu = \frac{\sigma}{ne}$$

$$= \frac{e}{h}\frac{1}{2\alpha^2}\frac{k_F}{n_{imp}}\frac{1}{H\!\left(\sqrt{\dfrac{g\alpha}{2\pi}}\right)}. \tag{22}$$

Substituting equation of $n_{imp}$ for $E_F \gg E_{rms}$ (eq. 13), we get

$$\mu \approx \frac{e\hbar}{\pi}\left(\frac{v_F}{E_{rms}}\right)^2\frac{J\!\left(\sqrt{\dfrac{g\alpha}{2\pi}}k_F\right)}{H\!\left(\sqrt{\dfrac{g\alpha}{2\pi}}\right)} \tag{23}$$

### 3.1 Thomas– Fermi

Within Thomas-Fermi approximation, we can obtain analytical expression for $H(z)$ [5-7]

$$H_{\mathrm{TF}}(z) = \int_0^1 d\eta \, \frac{\eta^3(1-\eta^2)}{\left(\eta^2+z^2\right)^2}$$

$$= \left(z^2+\frac{1}{2}\right)\ln\left(1+\frac{1}{z^2}\right)-1 \qquad (24)$$

Hence, the mobility for $E_F \gg E_{\mathrm{ms}}$ is given by

$$\mu \approx e\hbar \sqrt{\frac{\pi}{8g\alpha}} \left(\frac{v_F}{E_{\mathrm{ms}}}\right)^2 \frac{1}{H_{\mathrm{TF}}\left(\sqrt{\frac{g\alpha}{2\pi}}\right)} \qquad (25)$$

### 3.2 Random Phase Approximation

RPA mobility for $E_F \gg E_{\mathrm{ms}}$ is given by similar expression

$$\mu \approx \frac{e\hbar}{\pi}\left(\frac{v_F}{E_{rms}}\right)^2 \frac{J_{RPA}\left(\sqrt{\frac{g\alpha}{2\pi}}k_F\right)}{H_{RPA}\left(\sqrt{\frac{g\alpha}{2\pi}}\right)} \qquad (26)$$

$$H_{\mathrm{RPA}}(z) = \int_0^1 d\eta \, \frac{\eta^3(1-\eta^2)}{\left[\eta^2+z^2\tilde{\Pi}_{\mathrm{RPA}}(2k_F\eta)\right]^2}, \qquad (27)$$

where both $J_{\mathrm{RPA}}$ $H_{\mathrm{RPA}}$ need to be evaluated numerically. Both Thomas– Fermi and RPA upper and lower bound for mobility for region A, B, and C are computed in table S2.

TABLE S2: Mobility for $E_F \gg E_{\mathrm{ms}}$ calculated using Thomas– Fermi and RPA for both region A, B, and C.

|  | Region A | | Region B | | Region C | |
|---|---|---|---|---|---|---|
|  | $\alpha=0.069$ | $\alpha=0.174$ | $\alpha=0.069$ | $\alpha=0.174$ | $\alpha=0.069$ | $\alpha=0.174$ |
| $\mu^{\mathrm{TF}}\left(\mathrm{m^2V^{-1}s^{-1}}\right)$ | 1.7 | 0.6 | 3.0 | 1.1 | 5.2 | 1.9 |
| $\mu^{\mathrm{RPA}}\left(\mathrm{m^2V^{-1}s^{-1}}\right)$ | 1.7 | 0.6 | 3.0 | 1.1 | 5.2 | 1.9 |

Here, we have used $g=4$ to take into account valley and spin degeneracy and ultraviolet cutoff $\Delta/(2k_F)=100$ for RPA. We can see that both Thomas– Fermi approximation and RPA yield the same mobility.

**S6. DFT calculations of Na vacancies in the Na₃Bi lattice**

We have used Density Functional Theory (DFT) calculations to investigate the electronic properties of Na₃Bi slabs with Na-vacancies at the surface. There are two inequivalent Na atoms in the unit cell: Na(1) atoms [represented by dark blue spheres in Fig. 3(b)]; and Na(2) atoms [represented by the light blue and yellow spheres in Fig. 3(b)]. Total energy calculations indicate that the Na(2) surface vacancies [i.e. absence of the yellow atom in Fig. 3(b)] are energetically more favorable than Na(1) surface vacancies.

As can be seen in the leftmost bandstructure of Fig. 3(d), the introduction of a Na(2) vacancy in the bulk Na₃Bi slabs (2x2x2 supercell) preserves the gapless Dirac semimetal structure of pristine Na₃Bi. Nevertheless, as can be seen in panels (a) and (c) of Fig. S4, the position of the Dirac point is slightly moved away from the Gamma-point, while the lowest energy bands become flatter in the vicinity of the Dirac point. The removal of a Na(2) atom shifts the Fermi level to the valence band, as expected for the removal of this low electronegativity atom. Integrating the DOS in Fig. S5 for the thin-slabs from the Fermi level to the band gap gives one hole per Na(2) vacancy. Equivalently, we note that the removal of a single Na(2) atom removes 9 valence electrons from the system, but only 8 spin non-degenerate bands below $E_F$. This indicates that the Na(2) vacancy can be seen as effectively accepting one electron. Furthermore, there is no defect-induced flat band or localized state in the bandstructure, nor a resonance in $D(E)$ as shown by the full black curve in Fig. S5.

Confinement of the electrons along the z-direction by considering a finite slab [along the vertical direction in Fig. 3(b)] a bandgap is opened (with its size dependent on the thickness of the Na₃Bi as expected).[8] This gap is significantly larger in the slab with a Na(2) vacancy at the surface, when compared with the gap of a slab without Na-vacancies – see panels (b) and (e) of Fig. S4 corresponding to bilayers of Na₃Bi, respectively, without and with Na(2) vacancy at the surface. Nevertheless, as expected, increasing the thickness of the Na₃Bi slab with a Na(2) vacancy at the surface decreases the magnitude of the gap – compare panels (e) and (f) of Fig. S4 where we show the bandstructure of a Na₃Bi bilayer slab and of a Na₃Bi trilayer slab, both with a Na(2) vacancy (2x2 supercell).

The slabs with Na(2) vacancies at the surface show pronounced peaks in the $D(E)$ near the gap edges – see green and red curves in Fig. S5. Such resonances are not present in the $D(E)$ of either the pristine slab (i.e., slab without vacancy) or the slab with an interior Na(2) vacancy, as shown respectively by the dashed-black and the full-blue curves of Fig. S5. Notably, the trilayer slab with a Na(2) surface vacancy, shows not only a narrower bandgap, but also a more pronounced valence band resonance than the conduction band resonance. This result is consistent with the experiment, where no $D(E)$ resonance is seen at the conduction band. Furthermore, the ab initio $D(E)$ peaks appear at approximately the same energy position (~50meV below the band edge for the trilayer) of the resonant feature experimentally observed (~35meV below $E_D$).

As can be seen in Fig. S4(e,f), the peaks in the $D(E)$ result from the formation of a "Mexican hat" shaped valence and conduction band edges in the bandstructure. The fact that by increasing the slab thickness (from bilayer to trilayer) the $D(E)$ peaks magnitude changes only slightly, while the confinement gap decreases sharply (from 1.39 eV to 0.70 eV), suggests that the peaks do not scale with the size of the confinement gap, but rather originate from the presence of a surface. This is further confirmed from the absence of a resonant peak (as well as absence of "Mexican hat" in the bandstructure) when the Na(2) vacancy is in the middle of the 2x2x2 slab – see Fig. S4(d) and the full-blue line in Fig. S5.

From the DFT calculations we can then conclude that the observed resonance in the *D(E)* in the experiment is a result of the modification of the bandstructure due to the presence of the surface Na(2) defect, coupled with the formation of a confined state.

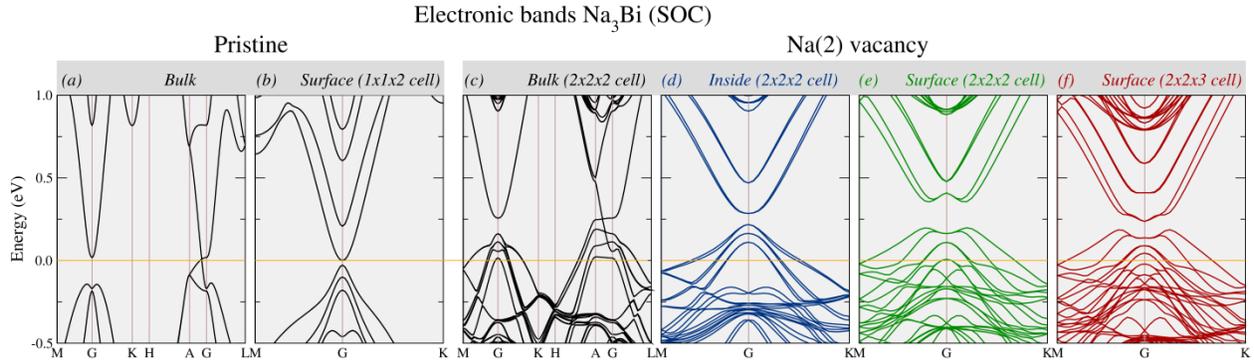

**Figure S4. Electronic structure of pristine and defective Na₃Bi.** (A) Pristine bulk. (B) Pristine bilayer simulated with a 1x1 supercell. (C) Bulk with a Na(2) vacancy simulated with a 2x2x2 supercell. (D) Bilayer with an interior Na(2) vacancy simulated with a 2x2 supercell. (E) Na(2) vacancy at the surface of a bilayer (simulated with a 2x2 supercell). (F) Na(2) vacancy at the surface of a trilayer (simulated with a 2x2 supercell). All electronic dispersions were calculated including SOC effects.

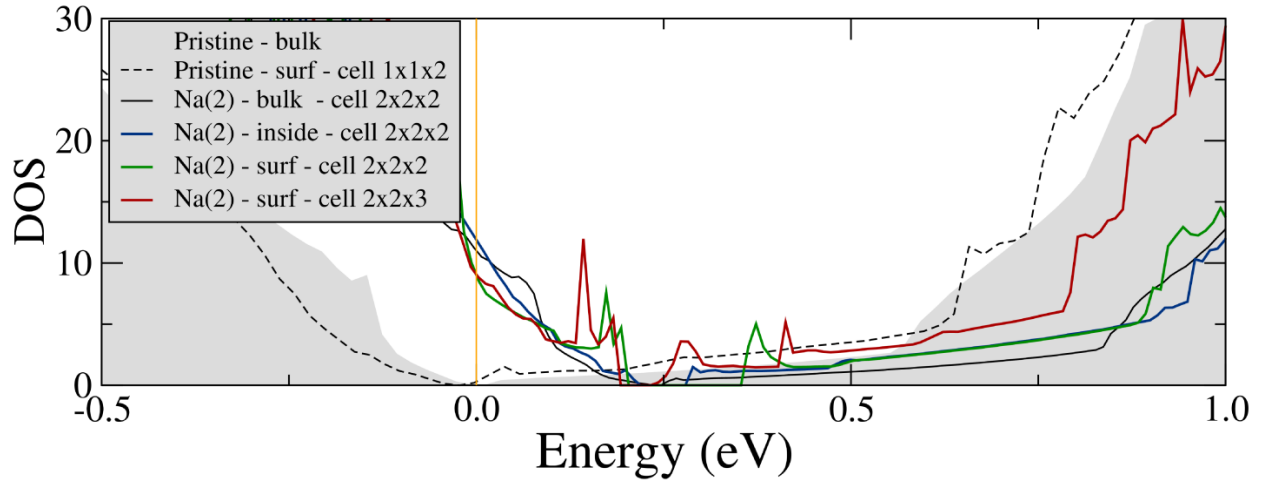

**Figure S5. Total density of states (DOS) of pristine and defective Na₃Bi.** The grey shadow stands for the DOS of the pristine bulk Na₃Bi, while the dashed-black curve stands for the DOS of the pristine bilayer slab. The black-full curve stands for the DOS of the bulk Na₃Bi with a Na(2) vacancy (2x2x2 supercell), the blue curve identifies the DOS of a bilayer slab with an interior Na(2) vacancy (2x2 supercell), while the green and red curves stand for the DOS of, respectively, a bilayer and a trilayer slab with a surface Na(2) vacancy (2x2 supercell).